\documentclass{aac}

\usepackage[numbers]{natbib}
\usepackage{rotating}
\usepackage{graphicx}

\usepackage{xcolor} 

\begin{document}

\title{Multileaf Collimator for Real-Time Beam Shaping using Emittance Exchange}

\author[aff1]{Nathan Majernik\corref{cor1}}
\author[aff1]{Gerard Andonian}
\author[aff2]{Ryan Roussel}
\author[aff3]{Scott Doran}
\author[aff3]{Gwanghui Ha}
\author[aff3]{John Power}
\author[aff3]{Eric Wisniewski}
\author[aff1]{James Rosenzweig}

\affil[aff1]{Department of Physics and Astronomy, University of California Los Angeles, California 90095, USA}
\affil[aff2]{Department of Physics, University of Chicago, Chicago, Illinois 60637, USA}
\affil[aff3]{Argonne National Laboratory, Lemont, Illinois 60439, USA}
\corresp[cor1]{NMajernik@g.ucla.edu}

\maketitle

\begin{abstract}
Emittance exchange beamlines employ  transverse masks to create drive and witness beams of variable longitudinal profile and bunch spacing. Recently, this approach has been used to create advanced driver profiles and demonstrate record-breaking plasma wakefield transformer ratios [Roussel, R., et al., Phys. Rev. Lett.  {\bf 124}, 044802 (2020)], a crucial advancement for efficient witness acceleration. However, since the transverse masks are individually laser cut and installed into the UHV beamline, refinement of the beam profiles is not possible without replacing masks. Instead, this work proposes the use of a UHV compatible multileaf collimator as a beam mask. Such a device permits real-time adjustment of the electron distribution, permitting greater refinement in a manner highly synergistic with machine learning. Beam dynamics simulations have shown that a practically realizable multileaf collimator can offer resolution that is functionally equivalent to that offered by laser cut masks.
\end{abstract}

\section{INTRODUCTION}
In the context of beam-driven wakefield accelerators, \emph{transformer ratio} is the ratio of the maximum accelerating field experienced by a witness bunch to the maximum decelerating field experienced by the driver bunch: $\mathcal{R} \equiv |W_+/W_-|$ \cite{RousselThesis}. Regardless of the peak accelerating gradient achieved in the wakefield accelerator, the transformer ratio sets an upper bound on the maximum energy gained by the witness beam before the driver has been depleted. For temporally symmetric drive bunches, the fundamental theorem of beam loading indicates that the transformer ratio cannot exceed 2 \cite{bane1985collinear}. However, by using a drive bunch with a tailored current profile it is possible to achieve values of $\mathcal{R}$ which exceed 2 \cite{Loisch:prl}, thereby reducing the required drive beam energy to accelerate a witness by a fixed, target energy. This has been demonstrated experimentally, including by \cite{gao2018observation} which measured a transformer ratio of 4.8 in a dielectric wakefield accelerator, and by \cite{roussel2020PRL} which measured a record-setting transformer ratio of 7.8 in a plasma wakefield accelerator (PWFA). Both of these examples relied on highly asymmetric drive beams to more effectively couple energy from the drive beam to the witness beam.

A number of options have been established for creating shaped current profiles, including laser pulse stacking \cite{loisch2018photocathode}, combining wakefield chirping with chicanes \cite{andonian2017generation}, the use of doglegs and higher order multipole magnets \cite{england2008generation}, as well as emittance exchange beamlines \cite{gao2018observation, roussel2020PRL}. Each technique has costs and benefits but emittance exchange (EEX) is one of the most versatile options for creating high charge bunches with a great degree of control. EEX is an advanced phase space manipulation technique whereby the transverse phase space of a beam is exchanged with its longitudinal phase space; one method of achieving EEX involves placing a transverse deflecting cavity between two doglegs \cite{RousselThesis}. By masking the beam’s transverse profile prior to EEX, high charge current profiles can be realized that would be difficult or impossible to achieve using other longitudinal shaping techniques. The EEX beamline at Argonne's AWA facility \cite{Ha:2017prl} was used to produce the record-setting transformer ratio of \cite{roussel2020PRL}.

Presently, at AWA's EEX beamline the transverse masking is done using laser-cut tungsten masks. Therefore, to change the mask shape requires installing newly cut masks in to the UHV beamline, an operation which takes days from start to finish. This latency makes it difficult to quickly iterate and refine the current profile to optimize a wakefield interaction. This current work proposes to replace these laser cut masks with a multileaf collimator (MLC), a device with dozens of independently actuated leaves which move in and out of the beam to create a custom aperture \cite{jordan1994design,boyer1992clinical,ge2014toward}. 
The most common application for multileaf collimators is their use in radiotherapy, where they can be used to shape the beams to precisely match the shape of the tumor from any angle, reducing damage to nearby tissue while still delivering an effective radiation dose. In the context of an EEX beamline though, the MLC will enable real-time, nearly arbitrary control over both the drive and witness bunch profiles. A multileaf collimator has many free variables for tuning, making it highly synergistic with machine learning. 

\section{BEAM DYNAMICS SIMULATIONS}
\begin{figure}[h]
\centerline{\includegraphics[width=\linewidth]{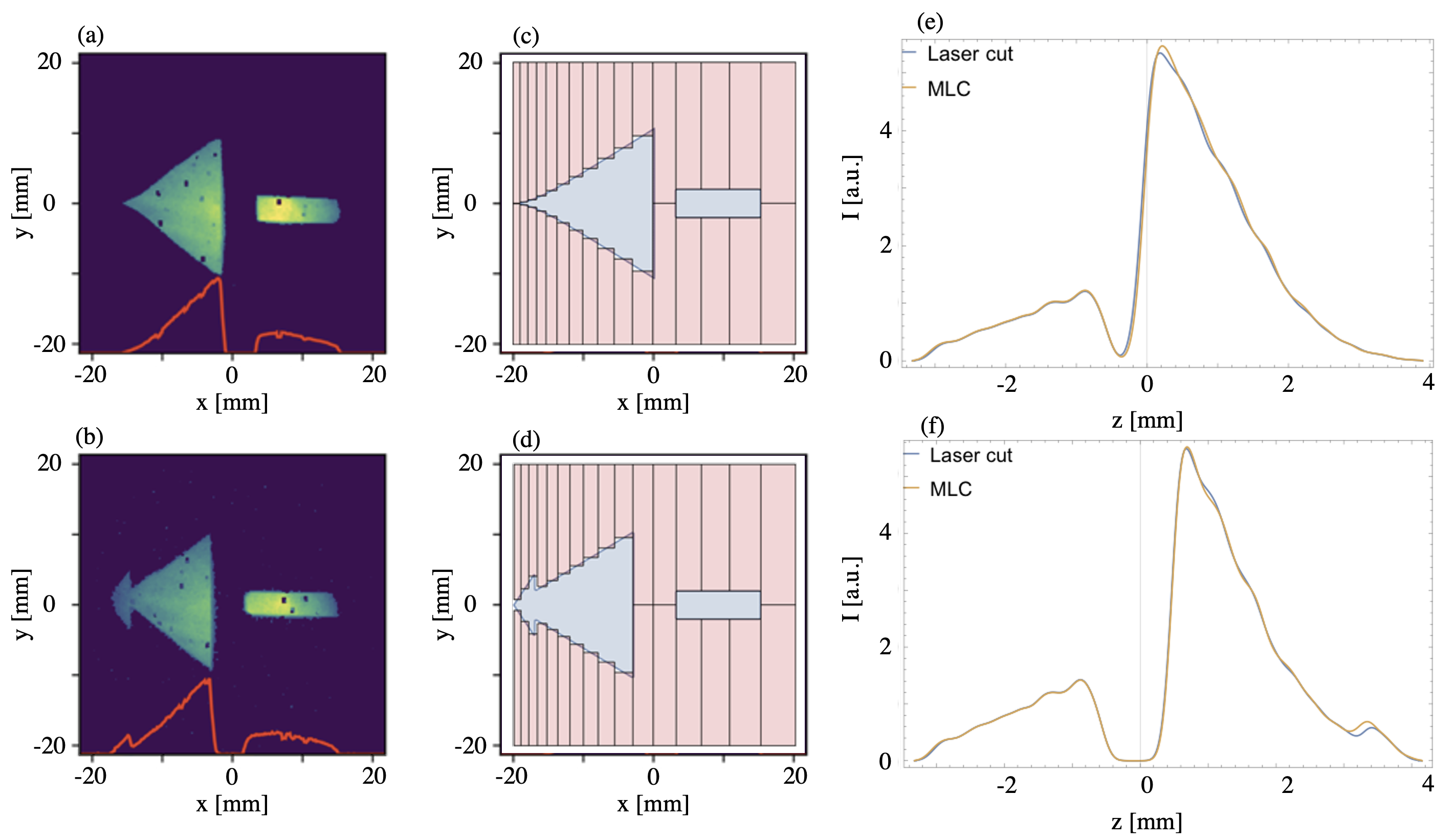}}
\caption{(a-b) The left column shows the YAG screen output shortly downstream from laser cut masks in the AWA EEX beamline from \cite{RousselThesis}; the red lineout is the result of integrating screen brightness in the vertical direction. (c-d) The results from the two laser cut masks are represented as the blue regions and the pink bars represent the MLC leaf positions to approximate this aperture. (e-f) For each row, the masked beams are virtually propagated through the EEX beamline with the final current profiles resulting from the target aperture (blue line) and the MLC approximation (orange line) shown.}
\label{fig:BeamDynamicsSims}
\end{figure}

Unlike a laser cut mask, a multileaf collimator is comprised of discrete elements, resulting in a `pixelation' of the masked beam. It is necessary to establish that, for a MLC with a practical number of leaves of an achievable width, relevant current profiles can be reproduced with sufficient fidelity to be useful for high transformer ratio wakefield acceleration.
To that end, beam dynamics simulations using OPAL \cite{adelmann2009object} have been conducted which compare the performance of laser cut masks to their approximations by a MLC. The simulated MLC has 32 leaves, 16 per side, with a logarithmic width distribution with an average leaf width of 2.5 mm. To test the MLC, two masks from \cite{roussel2020PRL} were selected; a YAG screen shortly downstream shows the masked beam in Fig. \ref{fig:BeamDynamicsSims}(a-b). The simulated MLC leaves were set to approximate this aperture in Fig. \ref{fig:BeamDynamicsSims}(c-d) before the masked results were virtually propagated through the EEX beamline to the plasma interaction point, with the current profiles for both cases for both the MLC and laser cut masks shown in Fig. \ref{fig:BeamDynamicsSims}(e-f). Based on the agreement between the MLC and laser cut mask current profiles, it is evident that a practical MLC is functionally equivalent to laser cut masks, yet provides greater flexibility for near-arbitrary beam shaping.

\section{CONCEPTUAL MODEL}

\begin{figure}[t]
\centerline{\includegraphics[width=\linewidth]{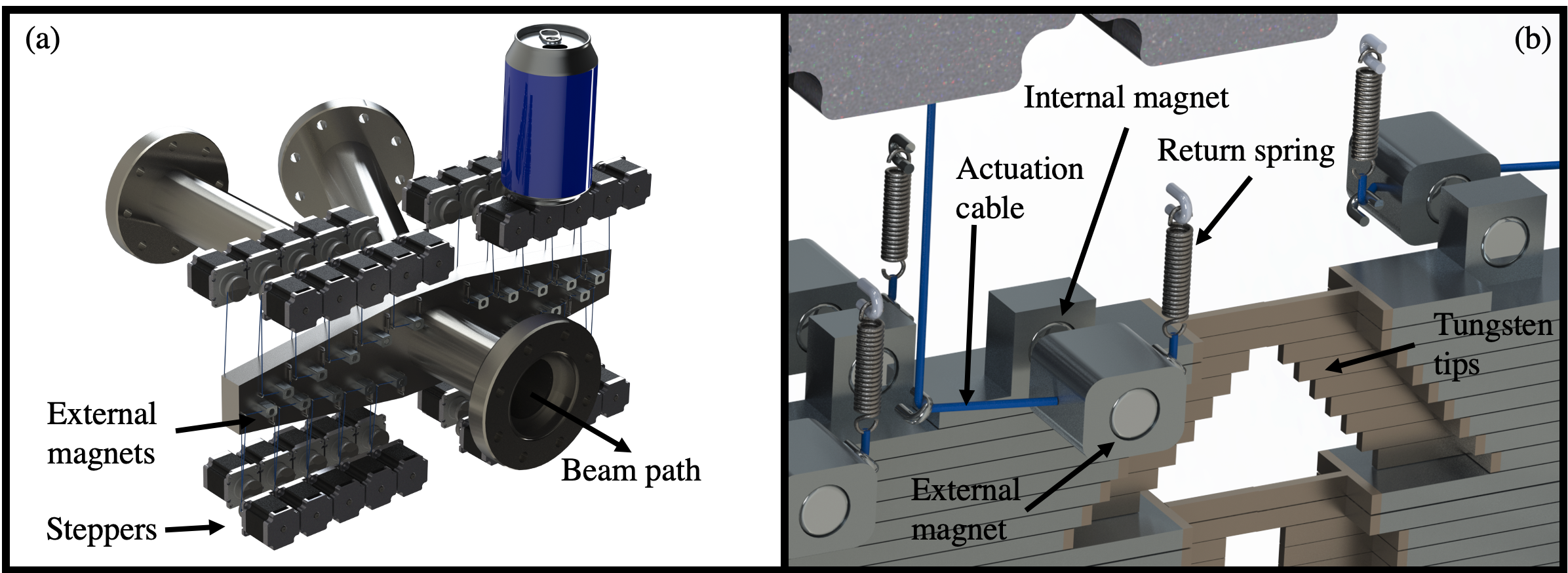}}
\caption{(a) A render of the proposed multileaf collimator with a standard 12 ounce soda can for scale. The ``wye'' beam pipe allows a camera to be pointed at the leaf tips for machine vision based feedback. (b) A detailed render of the MLC mechanism with the vacuum vessel not shown. Some of the key elements are called out.}
\label{fig:ConceptualModel}
\end{figure}

There are many differences between commercially available multileaf collimators for radiotherapy and the experimental requirements for EEX masking but the most significant is the need for UHV compatibility. In Fig. \ref{fig:ConceptualModel} a conceptual model which addresses these experimental requirements is presented. This design is a 40 leaf MLC, with 20 leaves per side, where each leaf is 2 mm wide. Each leaf is independently magnetically coupled to the exterior of the vacuum chamber and actuated via a cable driven by a stepper motor. The leaf is mostly comprised of aluminum to minimize weight but is tipped with tungsten for the beam interaction. Each leaf can move in and out by twelve millimeters, enough to cover the whole beam spot size. 

The magnetic coupling and cable actuation are a UHV safe and economical solution, but it comes with a substantial potential problem: the ability to dead reckon the positions of the leaves is lost since there is no rigid connection to a position encoder. Determining the leaf position therefore requires another form of feedback. To precisely read out the leaf positions the YAG screen images, like those of Fig. \ref{fig:BeamDynamicsSims}(a-b) can be used, but this measurement is destructive to the beam. To supplement this diagnostic, a camera will be pointed at the MLC leaves and, using machine vision, the leaf positions can be determined in a fully online, nondestructive fashion.

\section{PROOF-OF-CONCEPT EXPERIMENTS}
For this concept to work, the magnets on the exterior of the vacuum vessel must stay coupled to the magnets on the interior, but there is a tradeoff between compactness and coupling strength. The coupling strength needs to exceed the leaf weight, magnet-to-chamber friction, other friction sources, and have some safety margin. It is expected that the total force will be dominated by the friction at the magnet-chamber interface since the magnetic attraction will produce a substantial normal force.

\begin{figure}[h]
\centerline{\includegraphics[width=0.75\linewidth]{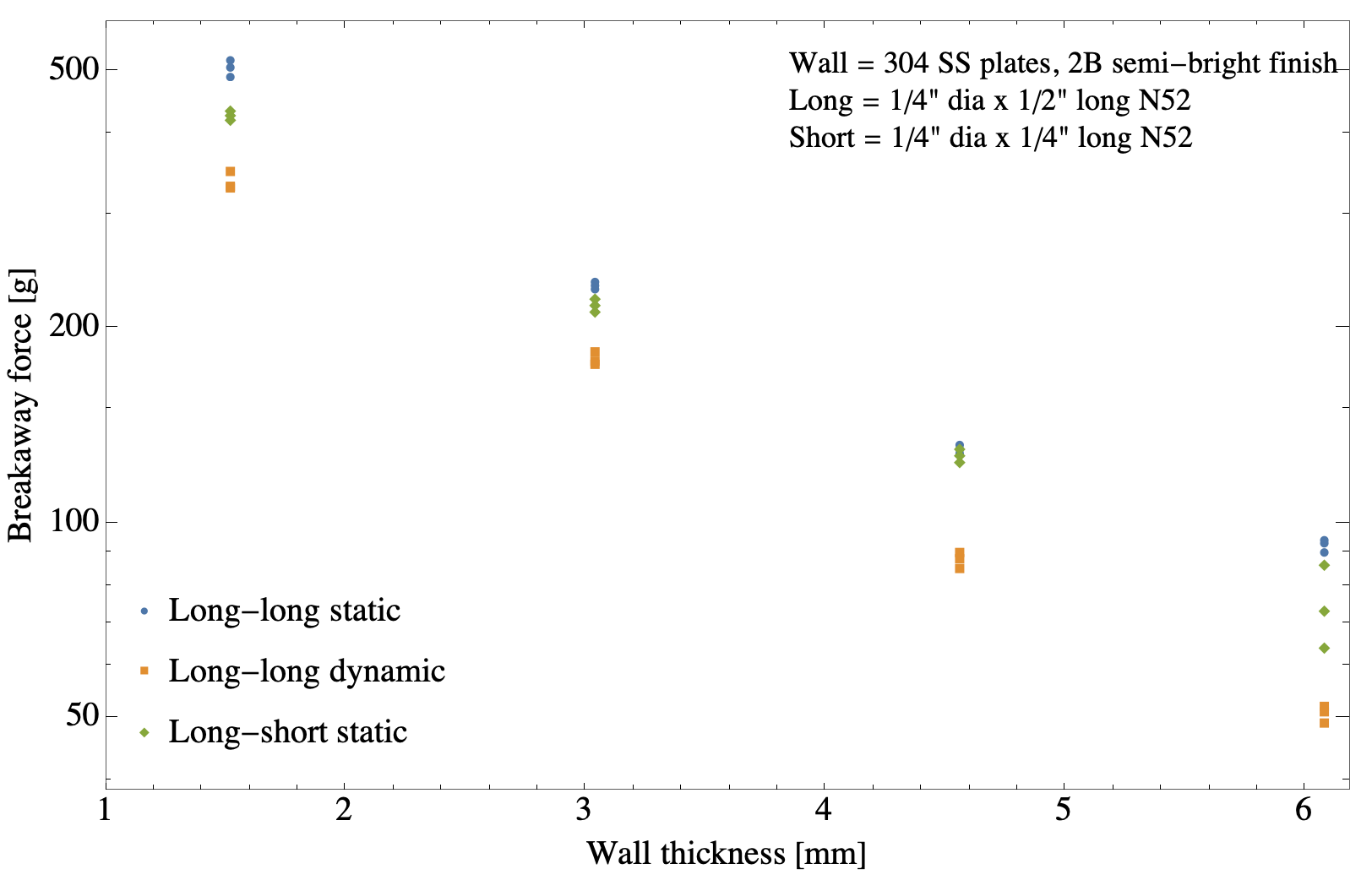}}
\caption{The results of the initial magnetic breakaway tests are shown. ``Long-long" tests uses 0.25 inch diameter, 0.5 inch long neodymium magnets for both the interior and exterior magnets while ``Long-short" tests use a 0.25 inch long magnet for the interior magnet. The ``static" tests represent the upper bound for breakaway strength in the limit of zero friction while the ``dynamic" case includes the friction of magnets on an unlubricated stainless steel surface.}
\label{fig:Friction}
\end{figure}

Preliminary tests were conducted using 0.25 inch diameter, N52 grade neodymium magnets (consistent with the 2 mm leaf design from Fig. \ref{fig:ConceptualModel}) for a variety of vacuum chamber surrogate wall thickness and magnet lengths to measure the force required to split the inner and outer magnet with the results shown in Fig. \ref{fig:Friction}. Two different test conditions were considered: static and dynamic. In the static cases, the wall surrogate and external magnet were held together and moved as a unit, therefore removing the magnet-chamber friction. Those results are representative of the breakaway strength anticipated if the inner and outer walls of the vacuum vessel were frictionless. In the dynamic cases, the external magnet was independently actuated and the breakaway force was again recorded. These cases, which include the motion of unlubricated magnets on a 2B semi-bright finish 304 stainless steel surface, breakaway at approximately 20 to 50 percent lower force, illustrating the important contribution of magnet-chamber friction. This performance can be recovered by reducing the friction but the vacuum requirements of the EEX beamline impose substantial restrictions on the options for lubrication. Overall though, the results suggest that the coupling strength is sufficient, even without lubrication. 

\section{SUMMARY AND FUTURE WORK}
A technology to replace the current laser cut tungsten masks used on the EEX beamline with a multileaf collimator has been presented. Beam dynamics simulations indicate that practically realizable multileaf collimator designs can produce current profiles that are functionally identical to laser cut masks. Such a change will allow real-time control over driver and witness current profiles allowing for iterative refinement that is not possible with a fixed-mask system. The large number of new variables the MLC introduces makes it highly synergistic with machine learning for the optimization of beam shaping for applications in high transformer wakefield acceleration. The conceptual design addresses the main hurdles of implementation and UHV compatibility. A series of benchtop tests are underway to validate the magnetic coupling and cable driven actuation concepts. A successful demonstration of the MLC concept would find utility in other accelerator beamlines that rely on transverse masking and require strict UHV levels, for example at BNL’s ATF \cite{BarberThesis} or at SLAC FACET \cite{Yakimenko}. 

\section{ACKNOWLEDGMENTS}
This work was supported by the National Science Foundation under Grant No. PHY-1549132 and DOE Grant No. DE-SC0017648.

\bibliographystyle{aac}%
\bibliography{aac2020_latex}%

\end{document}